\documentclass[aip,amsmath,amssymb,reprint, twocolumn]{revtex4-1}
\usepackage[utf8]{inputenc}
\usepackage[T1]{fontenc}
\usepackage{mathptmx}
\usepackage{rotating}
\usepackage{longtable}
\usepackage{listings}
\usepackage{xcolor}
\usepackage{siunitx}
\usepackage{lipsum}
\usepackage{wrapfig}
\definecolor{codegreen}{rgb}{0,0.6,0}
\definecolor{codegray}{rgb}{0.5,0.5,0.5}
\definecolor{codepurple}{rgb}{0.58,0,0.82}
\definecolor{backcolour}{rgb}{0.95,0.95,0.92}
\lstdefinestyle{mystyle}{
    backgroundcolor=\color{backcolour},   
    commentstyle=\color{codegreen},
    keywordstyle=\color{magenta},
    numberstyle=\tiny\color{codegray},
    stringstyle=\color{codepurple},
    basicstyle=\ttfamily\footnotesize,
    breakatwhitespace=false,         
    breaklines=true,                 
    captionpos=b,                    
    keepspaces=true,                 
    numbersep=5pt,                  
    showspaces=false,                
    showstringspaces=false,
    showtabs=false,                  
    tabsize=2
}

\usepackage{times,amsmath}
\usepackage{epsfig}
\usepackage{color}
\usepackage{longtable}
\usepackage{ulem}
\usepackage{graphicx}
\usepackage{dcolumn}
\usepackage{bm}
\usepackage{bookmark}
\usepackage{tabularx}
\usepackage{hyperref}
\usepackage{multirow}

\usepackage{tocloft}
\hypersetup{colorlinks=true, citecolor=blue, filecolor=blue, linkcolor=blue, urlcolor=blue}

\usepackage[switch, pagewise]{lineno}
\linenumbers\relax 
\nolinenumbers

\begin{document}

\title{Automated High-throughput Organic Crystal Structure Prediction via Population-based Sampling}

\author{Qiang Zhu}
\email{qzhu8@uncc.edu}
\affiliation{Department of Mechanical Engineering and Engineering Science, University of North Carolina at Charlotte, Charlotte, NC 28223, USA}

\author{Shinnosuke Hattori}
\email{shinnosuke.hattori@sony.com}
\affiliation{Advanced Research Laboratory, Research Platform, Sony Group Corporation, 4-14-1 Asahi-cho, Atsugi-shi 243-0014, Japan}

\date{\today}
\begin{abstract}
With advancements in computational molecular modeling and powerful structure search methods, it is now possible to systematically screen crystal structures for small organic molecules. In this context, we introduce the Python package High-throughput Organic Crystal Structure Prediction (\texttt{HTOCSP}), which enables the prediction and screening of crystal packing for small organic molecules in an automated, high-throughput manner. Specifically, we describe the workflow, which encompasses molecular analysis, force field generation, and crystal generation and sampling, all within customized constraints based on user input. We demonstrate the application of \texttt{HTOCSP} by systematically screening organic crystals for 100 molecules using different sampling strategies and force field options. Furthermore, we analyze the benchmark results to understand the underlying factors that may influence the complexity of the crystal energy landscape. Finally, we discuss the current limitations of the package and potential future extensions.\\

\end{abstract}

\maketitle
\makeatletter


\section{INTRODUCTION}

Molecular solids, referred to substances consisting of discrete molecules that are held together by relatively weak intermolecular forces, have been extensively used in chemical \cite{lee2011crystal, zhuo2022organic, yang-ANIE-2017, liu2018polymorphism}, pharmaceutical \cite{Neumann-Ncomm-2015} and semiconductor industries \cite{kallmann1960bulk, li2020molecular, yu2019crystal}. In these fields, the development of new organic materials with targeted properties relies heavily on understanding and controlling intermolecular interactions within the crystal structure.

Nowadays, data-driven computer simulation has been playing an increasingly important role in materials development \cite{Jain-NRM-2016}. Specifically, computational high-throughput screening of the existing organic crystals  has become popular for the design of new materials with the improved physical properties \cite{fratini2017map, friederich2019toward, saeki2019high, nematiaram2021bright, stuke2020atomic, kunkel2019finding}. However, most of the screening work has been limited by the availability of experimentally resolved structures. It has been found crystal polymorphism can effectively modify the physical properties \cite{Diao-JACS-2014, corpinot2018practical}. Nevertheless, to experimentally synthesize new polymorphs and characterize its structural properties is rather time-demanding and cost-expensive. 
From a practical standpoint, having the capability to screen likely organic crystal formations before laboratory synthesis and characterization would be highly valuable \cite{yang2018large}. 

In the past decade, there has been rapid development in the community of crystal structure prediction (CSP) for small organic molecules \cite{lommerse2000test, motherwell2002crystal, day:2005:blindcryst_3, day:2009:blindcryst_4, bardwell:2011:5th_blind_test_short, Reilly-Acta-2016}. The idea of CSP is to predict a short list of stable or metastable crystal packing that are likely to be observed in the experiment through the powerful structure exploratory algorithms \cite{Oganov-NRM-2019, Price-CSR-2014, gavezzotti2021crystalline, zhu2023organic}. Practically speaking, conducting a successful organic crystal prediction requires the integration of several different computational pipelines ranging from force field generation, structure sampling and post-process energy ranking. Most practitioners rely on subscriptions to commercial licenses to run simulation and analyze results. Due to the license restriction, it is often hard to reproduce historically published results, even for an experienced researcher. 

In the field of CSP for inorganic materials, there have been plenty of code choices (e.g., USPEX \cite{Oganov-JCP-2006}, AIRSS \cite{Pickard-JPCM-2011}, CALYPSO \cite{Wang-PRB-2010}, XtalOpt \cite{Lonie-CPC-2011}, GASP \cite{GASP-Python}) that are either completely open source or free for academic researchers. 
The availability of these tools has enabled significant progress in the field, allowing researchers to explore complex systems and improve their understanding of inorganic material behavior at a fundamental level \cite{Oganov-NRM-2019}.
While the field of CSP has been progressing rapidly, the code choices \cite{Oganov-JCP-2006, Case-JCTC-2016, Curtis-JCTC-2018} for the prediction of organic materials are still limited. More importantly, to the best of our knowledge, there is currently no open-source CSP code specifically designed for the high-throughput prediction of organic crystals. As organic crystals are increasingly relevant in fields like pharmaceuticals, organic electronics, and molecular materials, the development of such tools is becoming ever more critical to support the growing demand for rapid and accurate crystal structure prediction.

To promote the open-source activity in organic CSP, we have developed an open source code High-throughput Organic Crystal Structure Prediction (\texttt{HTOCSP}) that allows the automated prediction of organic crystal structures with a minimal input, by leveraging several existing open-source infrastructures \cite{pyxtal-Python, ost-Python, rdkit, amber, openff2.0.0} in molecular modeling. 
In the following section, we will begin by detailing the workflow, covering molecular analysis, force field generation, and crystal sampling. This will be followed by a benchmark study on 100 molecules, utilizing different sampling strategies and force field options. Finally, we will analyze the test results and conclude with a discussion of the current limitations and potential future extensions.

\begin{figure*}[ht]
\includegraphics[width=1.00\textwidth]{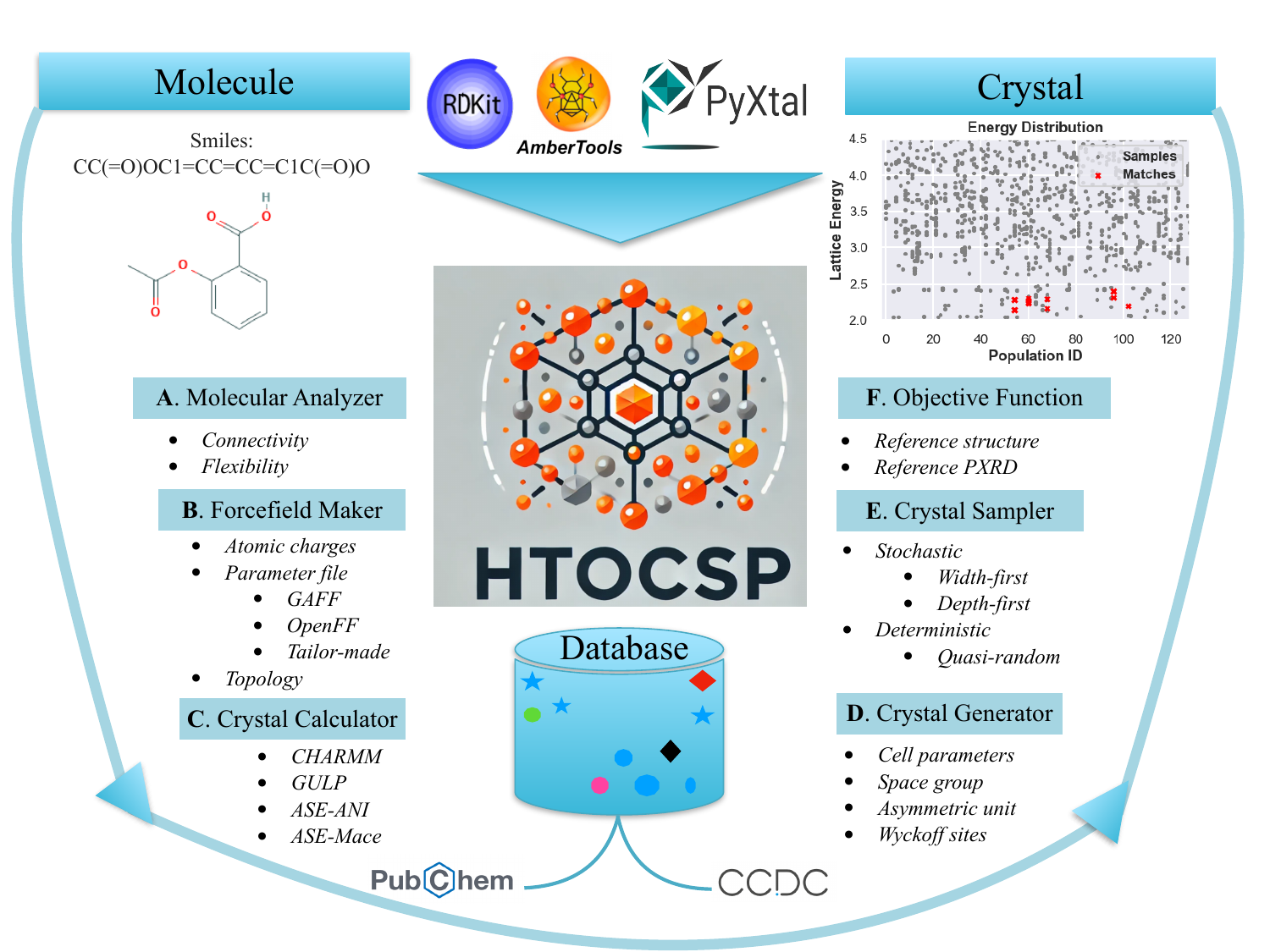}
\caption{\label{Fig1} The overview of HTOCSP workflow.}
\end{figure*}

\section{The HTOCSP Workflow}
In practical crystal structure prediction, the most common objective is to generate a shortlist of plausible crystal packings for a given molecular system. To address this challenge, multiple computational pipelines are often required to manage the system at both the molecular and crystal levels. In \texttt{HTOCSP}, we divide the entire crystal structure prediction process into six sequential tasks, as illustrated in Fig. \ref{Fig1}.

\subsection{Molecular Analyzer}
In organic chemistry, the most common way to represent a molecule is by using the SMILES (Simplified Molecular Input Line Entry System) string. SMILES strings use standard ASCII characters to denote atoms and bonds in a molecule. In \texttt{HTOCSP}, we utilize the \texttt{RDKit} library \cite{rdkit} to convert a SMILES string into 3D coordinates and analyze the flexible dihedral angles in the input molecule. For multicomponent systems, such as cocrystals, salts, and hydrates, each molecule must be handled separately.

In many cases, molecular SMILES strings and crystallographic information can be found through public online repositories such as PubChem \cite{pubchem} and the Cambridge Crystallographic Data Center (CCDC). However, it is important to note that some erroneous entries may exist. Therefore, an additional sanity check is strongly recommended before using this data.

\subsection{Force Field Maker}
Once the molecular information is known, we can proceed to extract the force field (FF) parameters. Currently, \texttt{HTOCSP} supports two main types of FF models based on the \texttt{AMBERTOOLS} \cite{amber} utility. The first type is the widely used General Amber Force Field (GAFF) \cite{wang2004development}, which covers C-H-O-N-S-P-F-Cl-Br-I elements to model small organic molecules. The second type is SMIRNOFF (SMIRKS Native Open Force Field), derived from the OpenFF initiative \cite{openff2.0.0}, which offers a more flexible and extensible way to describe force fields compared to traditional atom type-based methods. The OpenFF types support elements including C-H-O-N-S-P-F-Cl-Br-Li-Na-K-Rb-Cs. Due to its flexibility, it is also more convenient to extend support to other elements if needed. In addition, \texttt{AMBERTOOLS} is used to compute atomic partial charges using various schemes, such as Gasteiger, MMFF94, and AM1-BCC \cite{jakalian2000fast}. 

The extracted force field parameters, including descriptions of charges, bonds, angles, torsions, and nonbonded van der Waals interactions, can be saved as an Extensible Markup Language (XML) file according to the OpenFF standard. Optionally, companion topology files can also be generated in different formats for later use in structural relaxation codes. While we recommend the use of fixed GAFF/OpenFF parameters for most CSP applications, it is possible that the default GAFF or OpenFF parameters may poorly describe the system. In such cases, we allow users to retrain the parameters for better accuracy, similar to the tailor-made force field concept proposed by Neumann and coworkers \cite{Neumann-ANIE-2008} (see an extended discussion in Section \ref{challenges}). 

Finally, all the aforementioned functions have been implemented to the \texttt{pyocse} framework \cite{ost-Python} based on our earlier work \cite{Santos-Florez2023bending}. It is important to note that the current force field parameters were mainly fitted by data at standard temperature/pressure conditions. Therefore, it is not recommended to use it for the search of polymorphs at extreme conditions.

\subsection{Symmetry-constrained Structure Calculators}

With the classical force field in place, we are ready to build the structure calculator to perform geometry optimization and evaluate the energy of candidate crystals. Unlike ordinary geometry optimization tasks in molecular modeling, our goal is to optimize the crystal structure without breaking its symmetry or disrupting molecular connectivity \cite{QZhu-Acta-2012}. Currently, \texttt{HTOCSP} supports two symmetry-adapted molecular simulation codes suitable for CSP: \texttt{GULP} \cite{gulp} and \texttt{CHARMM} \cite{charmm}. In both codes, it is necessary to specify the molecular topology and symmetry operations, then instruct the calculator to optimize the cell parameters and molecular coordinates within the asymmetric unit. Optionally, the cell parameters can be fixed during optimization if they are already known.

In our earlier studies, \texttt{GULP} has been successfully used to address a range of practical CSP challenges \cite{QZhu-Acta-2012, Shtukenberg-CC-2017, Shtukenberg-CGD-2017, Tan-FD-2018, Yang-CGD-2019}. However, \texttt{CHARMM} generally offers faster performance than \texttt{GULP}, as it provides better support for the symmetry-adapted implementation of the Particle Mesh Ewald (PME) method \cite{PME}. Consequently, \texttt{CHARMM} is the default FF calculator choice in \texttt{HTOCSP} due to its faster computational speed.

Beyond classical force field models, there are ongoing efforts to develop generic or universal machine learning force fields for material and molecular screenings. Notable examples include ANI \cite{gao2020torchani} and MACE \cite{Batatia2022mace, Batatia2022Design}. These models have been trained comprehensively on extensive datasets, and the resulting parameters are available as open source for public access. Recent surveys \cite{zhu2023organic, kadan2023accelerated} indicate that machine learning force fields (MLFFs) generally reproduce experimental geometries more accurately and provide more reasonable rankings of crystal structures.

Despite these promising results, it’s important to note that most generic or universal MLFFs are trained primarily on known experimental structures and their nearby basins of attraction. Consequently, they may struggle with geometry relaxation if the initial structure is far from equilibrium or has a high energy. Due to these limitations, \texttt{HTOCSP} currently supports the use of ANI and MACE only for post-energy re-ranking on pre-optimized crystals generated by GAFF or OpenFF. These structure optimizations can be performed using a customized ASE calculator \cite{ase}, with the given space group symmetry constraints.

\subsection{Crystal Generator}
In a standard CSP task, the primary objective is often to search for plausible crystal packings within a shortlist of common space groups \cite{QZhu-Acta-2012, Price-CSR-2014, zhu2023organic}. Additional information, such as cell parameters pre-determined from experimental powder X-ray diffraction, can also be incorporated as input. This reduces the goal of CSP to generating multiple trial crystal packings within these constraints. To address this challenge, a structure engine is needed to generate random symmetric crystals.

Recently, we have developed a Python program called \texttt{PyXtal} that handles various structure generation and symmetry analysis tasks for 0D/1D/2D/3D atomic and molecular crystals \cite{pyxtal}. For the module dedicated to 3D molecular crystals, \texttt{PyXtal} enables the generation of trial structures with a customized asymmetric unit within a specified space group. In the asymmetric unit, one can place one or multiple molecules at the general Wyckoff positions. If the symmetry of the input molecule is compatible with the site symmetry of a given Wyckoff position, \texttt{PyXtal} also supports the assignment of the molecule to a special Wyckoff position with higher symmetry \cite{zhu2023organic}. This feature is particularly useful for generating crystal structures of high-symmetry molecules with a fractional $Z^\prime$ number (i.e., less than one molecule per asymmetric unit).

Additionally, it is quite common to encounter molecular crystal data labeled with a space group in non-standard settings. For instance, the standard setting for the most common molecular space group 14 is labeled as $P12_1$/$c1$ (shortened as $P2_1$/$c$). However, this space group can also be represented as $P12_1$/$n1$ or $P12_1$/$a1$, where the glide plane translates molecules in the diagonal direction or along the $a$-axis. Alternatively, one can choose the unique axis to be along $a$ ($P2_1$/$c11$) or $c$ ($P112_1$/$c$). Taking into account variations in space group notations due to the choice of origin, there are 530 concise space group settings according to Hall \cite{hall1981space}. In \texttt{PyXtal}, we allow the use of Hall notations to generate crystals in non-standard space group settings.

To illustrate the use of \texttt{PyXtal}, we provide a few examples in the following listing.

\begin{lstlisting}[language=Python, caption=PyXtal script to organic crystals], label={c1}]
from pyxtal import pyxtal
xtal = pyxtal(molecular=True)

# A random crystal in the 4e site of spg P21/c
xtal.from_random(3, 14, ['benzene'], [4], 
                 sites=[["4e"]])

# A random crystal in the 2a site of spg P21/c
xtal.from_random(3, 14, ["benzene"], [2], 
                 sites=[["2a"]])

# A random crystal in the `4e` site of spg P21/n
# P21/n correspond to Hall number 82
xtal.from_random(3, 82, ["benzene"], [4], 
                 sites=[["4e"]], 
                 use_hall=True)

# Random crystal from smiles
# Make sure rdkit is installed
mol = ["CC(=O)OC1=CC=CC=C1C(=O)O.smi"] # aspirin
xtal.from_random(3, 14, mol, [4], 
                 sites=[["4e"]])

\end{lstlisting}
 
\subsection{Population-based Sampling Methods}

After a random trial crystal is built, it needs to be relaxed to a configuration corresponding to a local energy minimum on the potential energy surface. By repeating this procedure many times, one can identify a complete set of plausible crystal packings that may exist in reality. Following this, their relative thermodynamic (and kinetic) stabilities must be evaluated using an accurate energy model.

With the aforementioned modules, one is ready to run a CSP calculation using a brute-force random search approach. This involves repeatedly generating random trial crystals and then using a structure calculator to perform geometry minimization, thereby obtaining low-energy structures. However, the success rate of finding the desired target may decrease exponentially as the search space becomes larger. 

To address this challenge, \texttt{HTOCSP} implements several population-based methods to achieve more efficient structure sampling. In global optimization, population-based methods refer to a class of optimization techniques that maintain and evolve a population of candidate solutions throughout the optimization process. Unlike traditional optimization methods that work with a single solution (e.g., gradient descent), population-based methods explore the search space using multiple solutions simultaneously. This approach helps to avoid local minima and increases the likelihood of finding a global optimum.

\begin{figure}[ht]
\includegraphics[width=0.5\textwidth]{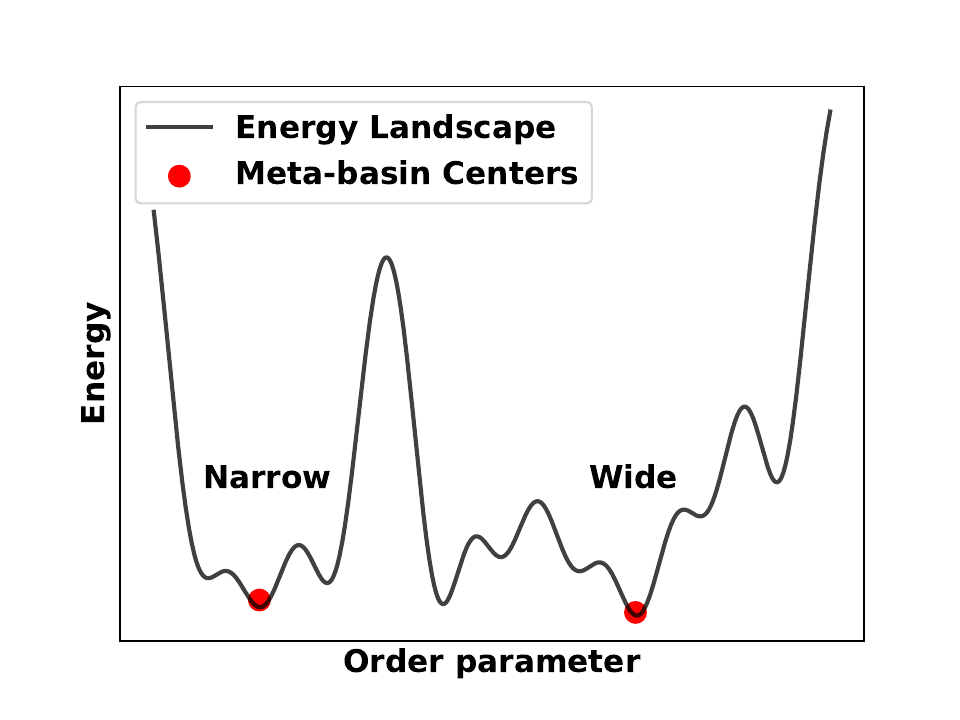}
\caption{\label{Fig2} The schematic representation of two meta-basins.}
\end{figure}

In the past, several population-based methods have been successfully applied to address challenges in both inorganic and organic crystal structure prediction, utilizing genetic/evolutionary algorithms \cite{Lyakhov-CPC-2013, GASP-Python, Lonie-CPC-2011, Curtis-JCTC-2018} and particle swarm optimization (PSO) \cite{Wang-PRB-2010}. These approaches often require mechanisms to generate new solutions by sharing information from multiple sources (e.g., the crossover operation in genetic algorithms or velocity updates via the Cognitive Social Component in PSO). However, handling such processes without breaking crystal symmetry is often challenging. Therefore, we decided to adopt a simplified population-based approach that avoids information sharing between individuals within the same generation during the evolution process.

Specifically, we begin with a population of randomly generated trial crystal structures and update each individual based on its own evolution history. For a given chemical, a large number of hypothetical structures can exist. After relaxation, their trajectories form basins of attraction on the potential energy surface. The probability of successfully finding a low-energy structure depends on the shapes and sizes of the hyper-volumes of these basins of attraction and the details of the search \cite{Pickard-JPCM-2011, Martiniani-PRE-2016, Ste-PRL-2016}. And we hypothesize that the target crystals (i.e., the likely observed crystal forms in experiments) form stable basins on the energy surface. However, due to artifacts in force fields and the constraints of symmetry and zero temperature during geometry optimization, these basins are often surrounded by many shallow energy minima. Consequently, a true basin and its surrounding shallow minima collectively form a meta-basin (as illustrated in Fig. \ref{Fig2}).

In a purely random search, there is a good chance of sampling within the meta-basin and landing in one of the surrounding shallow minima, but this may miss the true target structure (i.e., the center of the meta-basin) within a limited number of sampling attempts. Therefore, a wise approach is to perform a further local search to locate the centers of these meta-basins based on historically explored energy minima. Depending on the extent of effort in utilizing historical information, we devised two stochastic sampling strategies and one deterministic strategy as follows.

\subsubsection{Stochastic Width-First Sampling}
This approach resembles random sampling but with a structured twist \cite{Pickard-JPCM-2011}. Initially, all structures are randomly generated and then relaxed using the structure calculator. These structures are ranked by their final energies. In subsequent generations, a portion of the high-ranked (low-energy) structures undergo mutation by perturbing the cell parameters and molecular Wyckoff sites while preserving crystal symmetry. Conversely, the high-energy structures are discarded, and new random structures are generated in their place. The fraction of structures subject to mutation is a hyperparameter ranging between 0 and 1. When this parameter is set to 0, the method reverts to a pure random sampling algorithm.

Compared to pure random sampling, this method dedicates additional effort to exploring the surroundings of previously visited low-energy regions, thereby increasing the success rate of identifying more promising low-energy basins. Despite these extra efforts, the majority of the sampling still focuses on generating new random solutions, which is why we refer to this method as \textit{Width-First Sampling} (WFS). This type of sampling is particularly useful for detecting the target structure within a \textit{narrow meta-basin} with limited spread width, as illustrated in Fig. \ref{Fig2}.

\subsubsection{Stochastic Depth-First sampling} 
In a real system, the target structure may be located within a \textit{wide meta-basin} surrounded by numerous shallow minima (see Fig. \ref{Fig2}). In such cases, WFS may repeatedly visit some of these surrounding shallow minima without successfully identifying the true target state. To address this issue, we propose an alternative sampling strategy that prioritizes exploring the surroundings of already identified low-energy regions.

In this strategy, for each individual, we select either the most recently optimized structure (with a probability of 0.7) or the historically identified lowest-energy structure (with a probability of 0.3) to perform mutations and generate new samples. This process continues for 20 generations before restarting with new random samples. This mechanism is analogous to performing parallel basin hopping \cite{Wales-JPCA-1997} on multiple random samples. As a result, it achieves better sampling within each individual meta-basin, increasing the likelihood of identifying a true target surrounded by many shallow energy minima. Unlike WFS, which focuses on breadth, this approach is more focused on learning the shape of each meta-basin, which is why we refer to it as \textit{Depth-First Sampling} (DFS).

\subsubsection{Deterministic quasi-random sampling}
In the CSP community, the deterministic quasi-random sampling (QRS) method is also popular \cite{Case-JCTC-2016}. Quasi-random sequences are deterministic sequences of numbers designed to cover a multi-dimensional space more uniformly than uncorrelated random sequences. Unlike purely random sequences, which can result in clustering and gaps, quasi-random sequences aim to fill the space evenly, ensuring better coverage of the domain. Due to their deterministic nature, quasi-random sequences can be used to systematically cover the space and test the convergence of results without introducing randomness.

To perform QRS on molecular crystals, we consider cell parameters and molecular positions separately. For example, the aspirin crystal can be represented by the following variables.

\begin{table}[ht]
\caption{The representation of $P2_1$/$c$ aspirin form I in QRS.}\label{tab_1D}
\begin{tabular}{ll}
\hline
Hall Space Group Number:    & ~~\textit{81}                \\
Cell Parameters:            & ~~\textit{11.43,  6.49, 11.19, 83.31}  \\
Wyckoff Index:              & ~~4e \\
Molecular XYZ               & ~~\textit{0.77, 0.57, 0.53}\\
Molecular Orientation       & ~~\textit{48.55, 24.31, 145.9} \\
Molecular dihedral angles   & ~~\textit{-77.85, -4.40, 170.9} \\\hline
\end{tabular}
\end{table}

In the context of population-based methods, we generate a valid QRS sample for the cell parameters in each generation, followed by sampling the molecular positions, orientations, and dihedral angles with a sufficiently large population size. The user must define the lower and upper bounds for cell lengths to constrain the range of QRS samples, as well as specify a range of acceptable volume values to filter out unit cells that are either too large or too small.

\subsubsection{Termination Mechanisms}
Finally, all three sampling algorithms have been implemented in the \textit{pyxtal.optimize} module, and they can be conveniently accessed through \texttt{HTOCSP}. In practical CSP, we recommend a minimum population size of 256 in each generation. For the initial guess, the Bravis lattices are generated randomly within a range of allowed volumes. The crystal volumes ($V$) are estimated based on the molecular volumes and multiplicities. Then, two numerical factors (e.g., 0.7 and 1.3) are applied to define the lower and upper bound. The minimum and maximum vector lengths are applied based on its molecular shape and crystal symmetry, and the minimum and maximum angles are 30$^\circ$ and 150$^\circ$. All newly generated structures undergo multiple steps of geometry optimization using either GAFF or OpenFF. At the end of each optimization, it is possible that the unit cell may have very acute or obtuse angles, particularly in triclinic or monoclinic systems. In such cases, the angles are transformed to approach 90$^\circ$ whenever possible.

In principle, all sampling algorithms should run for a sufficiently large number of generations to ensure convergence. There are two mechanisms to terminate the sampling. If the sampling is run in \textit{production} mode (i.e., without knowledge of the target structure), it will stop once the maximum number of generations is reached. If the sampling is run in \textit{validation} mode with the target structure provided, the sampling may stop earlier as long as a good match with the target structure is detected 10 times.

To determine if a good matched structure is found, the StructureMatcher module in \texttt{Pymatgen} \cite{pymatgen-2013} is used. In this module, we ignore all H atoms and build a one-to-one map between each molecule in the unit cell, then check the largest root mean squared error (RMSE) between each atomic pair. By default, two structures are considered identical if the fractional length tolerance is 0.3, the site tolerance is less than 0.3 Å, and the angle tolerance is less than 5 degrees. It’s worth noting that Mercury \cite{mercury} also offers a mechanism to compute RMSE for 15-20 packed molecules via the COMPACK algorithm \cite{compack}. After extensive comparison, we found that these two functions produce nearly identical results. For ease of code implementation and accessibility, we have chosen to use \texttt{Pymatgen}’s StructureMatcher module in \texttt{HTOCSP}.

\subsection{Objective Functions}
In both WFS and DFS runs, the structures need to be ranked according to an objective function. By default, lattice energy is used as the optimization objective. However, it is also possible to choose an alternative objective function based on a similarity measure with respect to a reference powder X-ray diffraction (PXRD) pattern. To use this feature, users simply need to provide the experimentally measured PXRD raw data. During the search process, each structure will be used to compute a normalized similarity score ($S$) with the given reference PXRD using \texttt{PyXtal}, where a score of 1 indicates a perfect match and a score of 0 indicates no correlation. These $S$ values will be used for ranking the structures and will be recorded in the exported CIF files. This feature can guide the search toward the desired structure if the PXRD data is known. Currently, \texttt{HTOCSP} only supports the single objective optimization based on the PXRD match. Enabling the dual objective Pareto optimization \cite{Oganov-NRM-2019} is planned in the next release.

\section{Software Installation and Example Usage}
As described in the previous section, most of the structure generation and sampling algorithms have been implemented in \texttt{PyXtal}. However, a complete CSP calculation requires additional, more complex software dependencies, such as \texttt{RDKit}, \texttt{AmberTools}, and \texttt{CHARMM}. Therefore, we have integrated all these packages into the \texttt{HTOCSP} platform for the convenience of CSP practitioners.

For practical usage, one should first install the required Python packages via \textit{conda install}. Then, download and compile the source codes of \texttt{CHARMM} or \texttt{GULP}. The executable paths and necessary libraries must be added to the environmental variables.

To run the simulation, one typically needs to setup the following Python script.

\begin{lstlisting}[language=Python, caption=PyXtal script to run HTOCSP for the aspirin system.], label={c2}]
from pyxtal.optimize import WFS
from pyxtal.representation import representation

# Parameter Setup
sg, gen, pop, ncpu = [14], 10, 256, 128
wdir = 'aspirin-simple'
smiles = "CC(=O)OC1=CC=CC=C1C(=O)O"

# Obtain the reference crystal from 1d string
x = "81 11.38  6.48 11.24  96.9 1 0 0.23 0.43 0.03  -44.6   25.0   34.4  -76.6   -5.2  171.5 0"
rep = representation.from_string(x, [smiles])
xtal = rep.to_pyxtal()

# Sampling
go = WFS(smiles,
         wdir,
         sg,  # list of space groups
         tag = 'aspirin',
         N_gen = gen,
         N_pop = pop,
         N_cpu = ncpu,
         ff_style = 'openff')
go.run(ref_pmg=xtal.to_pymatgen())
go.print_matches(header='Ref-DFS-gaff')
go.plot_results()
\end{lstlisting}

\begin{figure}[ht]
\includegraphics[width=0.5\textwidth]{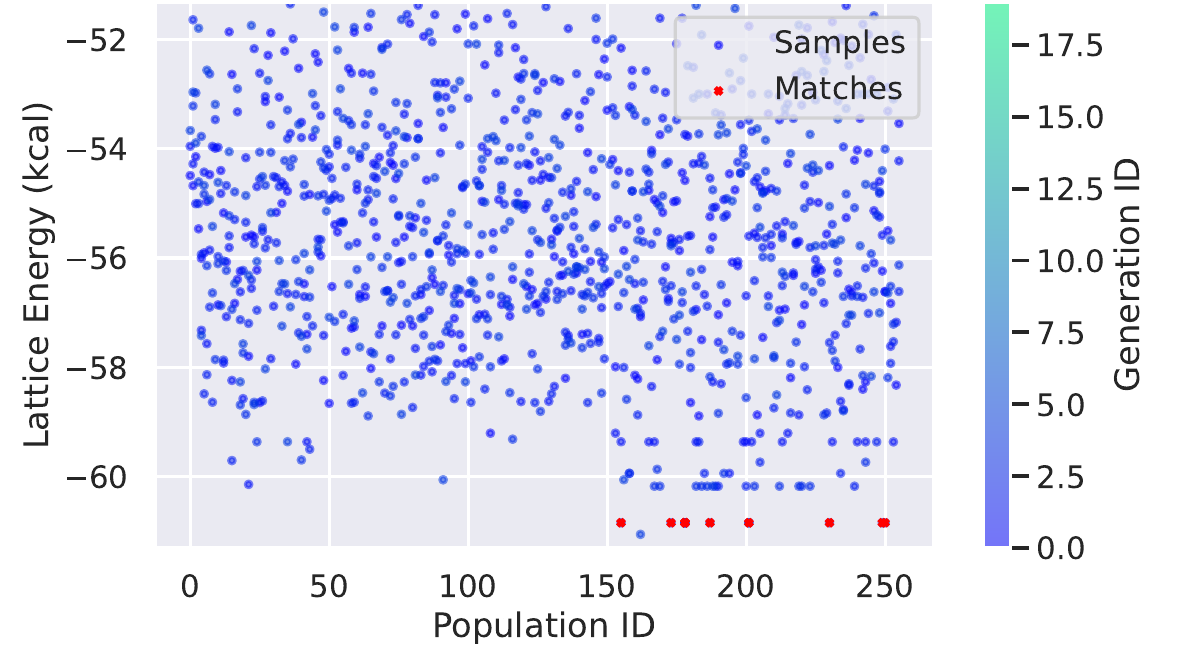}
\caption{\label{Fig3} The CSP results based on WFS and OpenFF for the aspirin crystal in space group 14 with $Z^\prime$=1 are presented in this plot. Each dot represents a sampled individual structure, denoted by a tuple of (generation ID, population ID, lattice energy). The matched structures are highlighted as red crosses for clarity.}
\end{figure}

This script will sample the aspirin crystals in \textit{WFS} mode using the OpenFF force field, with a population size of 256 over 20 generations. With 128 parallel cores, the total sampling time is approximately 2.5 generations per minute. Therefore, the entire run of 20 generations can be completed within 10 minutes. When compared to the experimental aspirin form I in space group 14 ($P2_1$/$c$) with $Z^\prime$=1, we found a total of 7 matches out of 5120 sampled structures, as shown in Fig. \ref{Fig3}. We also found another structure with lower energy corresponding to the form II of aspirin. These two forms have been well known due to very close lattice energy ranking and their relative stability can only be distinguished with advanced free energy calculation method \cite{hattori2024revisiting}. Last, it is important to note that the number of matched structures may vary due to the stochastic nature of the sampling algorithm. The confidence in these sampling results will be further discussed in the following section.

In a standard CSP run via \texttt{HTOCSP}, one only needs to provide a SMILES string to define the target molecule (either real or hypothetical), a list of space group choices, and a few sampling parameters related to population and generation sizes. If the target system is a cocrystal or salt, a list of SMILES strings and the compositional ratio must be provided. For further instructions, users can refer to the online documentation of \texttt{HTOCSP} \cite{HTOCSP} to learn how to set up other types of calculations, such as structure searches with respect to the reference PXRD data.

\section{Benchmark Results and Discussions}\label{bench}

In this section, we expand upon the example discussed in Fig. \ref{Fig3}. In a typical CSP calculation, the search space grows exponentially with the number of variables, including unit cell vectors, symmetry operations, molecular positions, conformations, orientations, and the number of molecules per asymmetric unit. Under blind test conditions, it is often difficult to be confident in the results obtained from limited sampling efforts. Clearly, some molecules may have more complex and fuzzy energy landscapes compared to others \cite{Price-CSR-2014}.

To validate the concept of energy landscape complexity, we performed a systematic benchmark test on a set of 100 experimentally reported crystals. We selected these systems to encompass a range of challenges encountered in realistic CSP simulations. The set includes 30 systems from past blind tests \cite{lommerse2000test, motherwell2002crystal, day:2005:blindcryst_3, day:2009:blindcryst_4, bardwell:2011:5th_blind_test_short, Reilly-Acta-2016}, 38 representative organic semiconducting (OSC) molecules, primarily consisting of thiophene rings, 14 polynuclear aromatic hydrocarbons (PAHs) \cite{desiraju1989crystal}, and 18 other systems based on our previous research for various purposes (Misc.). These systems vary widely, including single-component systems, multi-component cocrystals/salts, fractional to multiple $Z^\prime$, and from rigid molecules to complex molecules with multiple flexible dihedral angles. In addition, a large fraction of the structures are not the most stable polymorph but are metastable. A complete list of their associated Cambridge Structural Database (CSD) entries and basic descriptions is provided in Appendix Table \ref{s1}.

\begin{figure*}[htbp]
\includegraphics[width=1.0\textwidth]{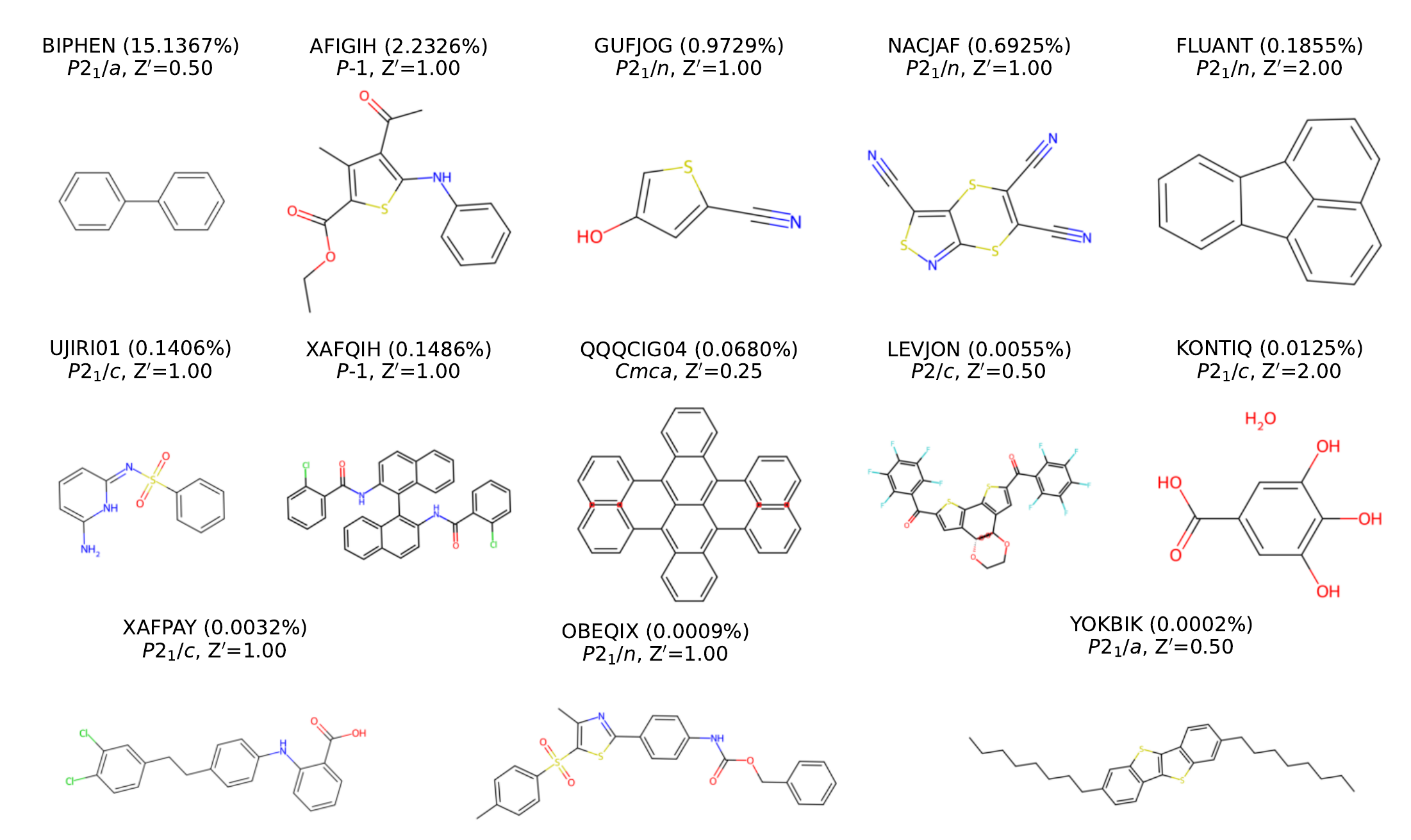}
\caption{\label{Fig4} The list of 13 representative molecules sorted by the sampling success rate values. The molecules are listed based on their entry names as deposited in the Cambridge Structural Database as shown in \url{https://www.ccdc.cam.ac.uk}.}
\end{figure*}

For each system, we performed four types of simulations: (1) WFS with GAFF (WFS-GAFF), (2) DFS with GAFF (DFS-GAFF), (3) WFS with GAFF relaxation and ANI energy evaluation (WFS-GAFF-ANI), and (4) DFS with GAFF relaxation and ANI energy evaluation (DFS-GAFF-ANI). In each run, we limited the search space based on the experimentally determined space group and $Z^\prime$. If the target crystal had a fractional $Z^\prime$ due to occupation of a special Wyckoff site, we followed the symmetry relation to transform it to a subgroup representation where the molecules occupy the general Wyckoff site \cite{pyxtal}. In all calculations, we set a population size of 256 with a maximum of 500 generations. 
However, the calculation could be terminated earlier if a total of at least 10 matched structures were found. In each simulation, we defined the success rate (SR) as the ratio of the number of matching structures to the total number of sampled structures. In each WFS run, the fraction of mutation was set to be 0.4.

\subsection{Categorization by Sampling Success Rate}
Fig. \ref{figs1} summarizes the whole results for all 100 systems. According to the averaged SR value of each system from four different sampling strategies, we divide them into four tiers as follows,

\begin{figure*}[htbp]
\includegraphics[width=1.0\textwidth]{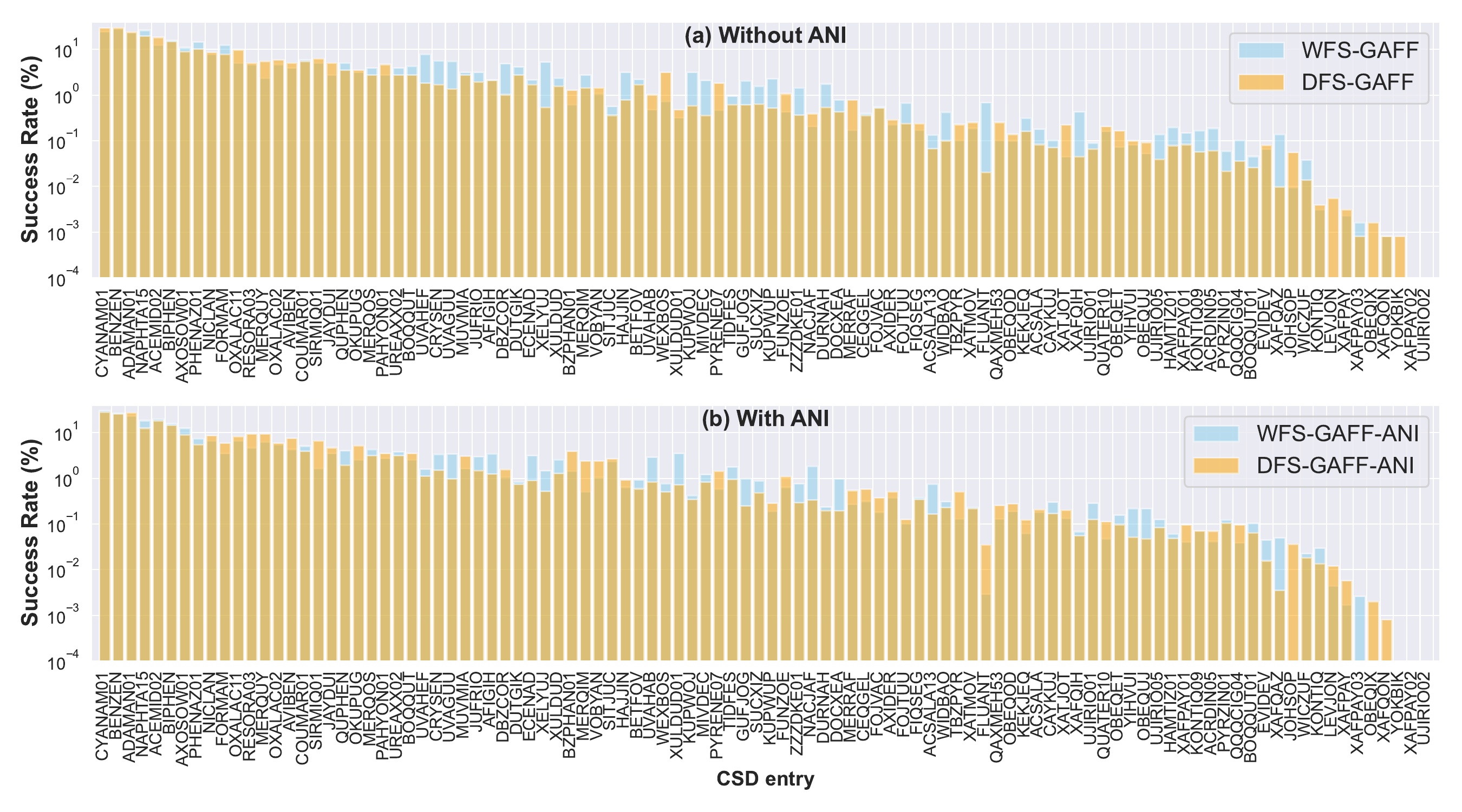}
\caption{\label{Fig5} The success rate distribution of 75 systems from four different sampling strategies.}
\end{figure*}

\begin{enumerate}
    \item \textbf{Tier I} (SR > 0.5\%). This tier consists of 56 systems that can be considered relatively easy CSP challenges. Statistically, one only needs to sample 2-3k structures to hit the target structure at least 10 times. At the molecular level, Tier I systems are mostly composed of symmetric molecules with low mass (< 300) and only a few flexible dihedral angles (< 3). Representative molecules in this tier include many PAH/OSC polymorphs (e.g., BENZEN: 27.5390\%; ADAMAN01: 24.3164\%; NAPHTA: 19.1406\%; UVAHEF: 3.0835\%; XELYUJ: 1.9458\%) and simple targets (e.g., AXOSOW01: 10.3516\%; BOQQUT: 3.2715\%; GUFJOG: 0.9729\%; NACJAF: 0.6925\%). At the crystal level, the $Z^\prime$ number is no more than 1. AFIGIH (2.2326\%) is an outlier in this group, as it has only $C1$ symmetry, a relatively high mass (303.4), and 6 flexible dihedral angles. However, as observed in other systems, crystal structures with $P$-1 symmetry are generally easier to predict than those with monoclinic or orthorhombic symmetries with more symmetry operation constraints.
    
    \item \textbf{Tier II} (0.05 \% < SR < 0.5\%). This tier includes 33 systems that present relatively moderate CSP challenges. Typical molecules in this tier include 14 blind test targets (e.g., WIDBAO: 0.2650\%; OBEQOD: 0.1755\%), two famous aspirin polymorphs (ACSALA13: 0.2777\%; ACSALA: 0.1617\%), and many OSC polymorphs. To ensure convergence of 10 hits for the target structure, up to 20-30k structural samples are required. Most systems in this group have either fairly flexible molecules (with 2-6 rotors) or more than one molecule in the asymmetric unit ($Z^\prime \geq$1). Notably, the blind test target XXVI XAFQIH, previously considered a challenging example, also falls into this group with a modest SR value of 0.1486\%.
    
    \item \textbf{Tier III} (0.001\% < SR < 0.05\%). This tier includes 6 systems (JOHSOP, WICZUF, KONTIQ, LEVJON, XAFPAY, and XAFPAY03). These systems typically require over 100k structural samples (approximately 4-6 hours with 128 parallel CPU processes). Systems in this tier are often characterized by either $Z^\prime$=2 or a high number of rotors (4-6). As we will discuss in the following, the sampling efficiency strongly depends on the choice of algorithms. An efficient algorithm can significantly speed up the finding of target structures. 
    
    \item \textbf{Tier IV} (SR < 0.001\%). This tier consists of 4 examples that represent the most challenging CSP cases. Despite running 128k samples with four different strategies, we observed only a few hits for OBEQIX (7), XAFQON (4), and YOKBIK (2), and we were unable to find the target structures of XAFPAY02 and UJIRIO02 within several runs of 500 generations. UJIRIO02 were found by a couple of times when the maximum number of generations is 3000. To reliably predict these systems, one either needs to spend significantly more CPU hours or to develop more efficient sampling algorithms are needed.
\end{enumerate}

In practical CSP applications, it is popular to search for plausible crystal packing with 10 to 15 common space groups \cite{zhu2023organic}. As a result, the calculated success rate may descrease by 10-20 times in read-world scenarios. For clarity, we also provide the molecular diagrams and the corresponding space group/$Z^\prime$ numbers for several representative systems from each tier in Fig. \ref{Fig4}. One may use the complete data in Table \ref{s1} as a test bed to develop a predictive model to infer the CSP landscape complexity purely from the molecular and crystal information for any arbitrary system. 

\subsection{Impacts of Sampling strategies}
We now proceed to analyze the impact of sampling strategy on the success rate. In Tier I, around 25 experimental targets, such as benzene (BENZEN), naphthalene (NAPHTA15), biphenyl (BIPHEN), and urea (UREAXX02), can be trivially found multiple times within just two generations of 512 structural samplings. For these systems, all four sampling strategies are expected to yield very similar statistical results. For clarity, we have excluded these systems from our main text and instead focus on analyzing the remaining 75 systems that require sampling over multiple generations, as shown in Fig. \ref{Fig5}.

Comparing Fig. \ref{Fig5}a and Fig. \ref{Fig5}b, we first observe that adding ANI energy evaluation does not improve the efficiency of finding the target structures for the majority of 100 systems. A few exceptional cases include KONTIQ, WICZUF, PYRZIN01, OBEQUJ, XAFQON. However, the inclusion of ANI adds about 50-100\% overhead to the calculation time. Therefore, it is recommended to run these samplings with pure GAFF or OpenFF to reduce computational costs.

Second, the comparison between WFS and DFS in Fig. \ref{Fig5}a shows that the success rates do not vary significantly for Tier I and most Tier II structures. For these structures, sampling at most 10,000 structures should be sufficient to obtain at least 10 hits of the target structure, ensuring successful sampling. However, there are a few exceptions in Tier II where WFS tends to outperform DFS. For example, in FLUANT (a rigid PAH molecule with $Z^\prime$=2 in $P2_1/n$) and QQQCIG04 (the semiconducting rubrene with $Z^\prime$=0.25 in $Cmca$), WFS can identify the targets about 5-10 times faster than DFS. This suggests that the experimental structures in these systems are characterized by relatively isolated, narrow meta-basins.

On the other hand, DFS tends to be more efficient for Tier III/IV systems. For instance, the success rate of DFS-GAFF for JOHSOP is 0.0551\%, which is significantly higher than the 0.0095\% achieved by WFS-GAFF. Additionally, DFS produces relatively decent SR values for LEVJON (0.0055\%), OBEQIX (0.0016\%), XAFQON (0.0008\%), and YOKBIK (0.0008\%), whereas WFS fails completely. These cases suggest that the target structures are likely characterized by wide meta-basins, making an in-depth exploration strategy more efficient.

Clearly, there is a trade-off between using WFS and DFS. In the future, more in-depth studies will be conducted to develop a predictive model that infers CSP complexity and further improves the success rate by enhancing sampling methods.

\section{Remaining Challenges}\label{challenges}
While the current \texttt{HTOCSP} provides a general framework for automated crystal sampling in a high-throughput manner, there is certainly room for further improvements. Below, we briefly discuss several immediate challenges.

\subsection{More Efficient Sampling Strategy}
Based on our past experiences, it appears advantageous to separate cell parameters from the rest of the crystal variables during sampling. If the cell parameters are known, the CSP task becomes significantly easier, often by several orders of magnitude. In early CSP studies by Desiraju and Gavezzotti \cite{desiraju1989crystal}, it was found that the crystal cell shapes of similar molecules could often be fitted using empirical relations. With the rapid progress in deep learning, it is reasonable to anticipate that predicting cell parameters from molecular and space group information could become achievable in the near future. Therefore, it may be worthwhile to bias structure sampling toward a reduced set of more promising cell parameters to further improve sampling efficiency.

Additionally, data-mining-based approaches could supplement the current \textit{ab initio} methods. In inorganic crystal prediction, chemical substitution on well-known prototypes is widely used to generate new candidate crystals. If the packing motifs of organic crystals can be effectively grouped \cite{zhu2022quantification}, one could analyze the occurrence of each packing prototype and use them as templates to generate new structures more efficiently.

\subsection{Uncertainty quantification}
Our benchmark results on 100 systems suggest a correlation between crystal landscape complexity and the given input molecular/crystal variables. In a standard CSP run, the goal is typically to generate as many trial crystal structures as possible. However, the development of organic materials may require consideration of many molecules, necessitating high-throughput polymorph screening within a reasonable time frame. Ideally, we want set a smaller number of generations for simple systems and more generations for molecules with complex landscapes. To achieve this, a more rigorous uncertainty model should be developed to automate the setup of the sampling efforts based on the input molecular and crystallographic information.

\subsection{CSP Post-analysis and iterative FF optimization}
So far, the primary focus of the current \texttt{HTOCSP} development has been on identifying the most promising candidate crystal forms through extensive structure sampling. However, a significant challenge remains in extracting the target structure from the sampled structure pool. As observed in Table \ref{s1}, it often requires sampling 1-100k structures to ensure success. Due to the limitations of force fields, it is quite possible that the true target structures have an unfavorable energy ranking at the force field level. This trend of energy misrank is likely to be more pronounced  when predicting the polymorphs at non-standard pressure and temperature conditions.
To address this challenge, one possible approach is to implement a multi-stage ranking process that refines the selection of candidate structures. This process could involve additional criteria, such as energy calculations using more accurate methods like DFT, to filter out less likely candidates. The development of a comprehensive post-analysis module will be the subject of our future work.

Additionally, one can start a massive CSP calculation with the generic FF, label the energy and forces for already sampled configuration with more accurate DFT or machine learning models, and then reparameterize the original force field parameters. Such a detour strategy has been proved successful in the past blind tests and many pharmaceutical applications \cite{Neumann-ANIE-2008, Neumann2008tailor, Neumann-Ncomm-2015}. The current \texttt{HTOCSP} begins to implement this function and it should be available in the near future.

\section{Conclusions}
In summary, we have presented \texttt{HTOCSP}, a comprehensive platform designed to facilitate high-throughput crystal structure prediction for organic molecules. By integrating various tools such as \texttt{PyXtal}, \texttt{Pyocse}, \texttt{RDKit}, \texttt{AmberTools}, and \texttt{CHARMM}, \texttt{HTOCSP} enables automated sampling and structure optimization, making it accessible for CSP practitioners to predict and analyze potential crystal forms efficiently.

Our systematic benchmark on 100 experimentally reported crystals demonstrates the effectiveness of different sampling strategies. We have shown that the choice of sampling strategy can significantly impact the success rate, especially for systems with varying degrees of landscape complexity. While Width-first sampling proves more effective for systems characterized by narrow meta-basins, the Depth-first sampling shows its strength in exploring wide meta-basins, particularly in more challenging CSP cases. In addition, these statistical results may serve as a ground for more efficient sampling methods development in the future.

Despite these advancements, challenges remain, particularly in improving the efficiency of sampling strategies and developing robust post-analysis tools to accurately identify target structures from large datasets. The current work lays the foundation for future enhancements, including the integration of machine learning techniques to predict cell parameters, the implementation of more rigorous uncertainty quantification methods, and the development of advanced post-analysis modules to refine structure ranking. And the continuous development of \texttt{HTOCSP} will focus on addressing these challenges, with the aim of further improving the reliability and efficiency of CSP simulations. 

\section*{Acknowledgments}
Q.Z. acknowledge the NSF (DMR-2410178) and Sony Group Corporation for their financial supports. The computing resources are provided by ACCESS (TG-MAT230046).

\section*{Data availability}
The \texttt{HTOCSP} source code, instructions, as well as scripts used in this study, are available in \url{https://github.com/MaterSim/HTOCSP}. The results from this work were obtained using the \texttt{HTOCSP}v-1.0 as released in 2024/10. The 100 crystal structures as discussed in this study can be directly inquired from \url{https://www.ccdc.cam.ac.uk} by searching their corresponding CSD entries (e.g., ACSALA).

\section*{AUTHOR CONTRIBUTIONS}
QZ proposed this idea. Both QZ and SH designed the research,
analyzed the calculations and wrote this manuscript.

\nolinenumbers

\section*{REFERENCES}
\bibliography{ref}

\clearpage
\renewcommand{\thetable}{A1}

\begin{longtable*}{l|c|c|c|r|r|c|c|c|c|l}
\caption{The summary of 100 molecular systems used in the benchmark test. The success rates are averaged from all four different sampling strategies, while the energy rank is obtained from the DFS-GAFF calculation as mentioned in Sec. \ref{bench}. The $T_\text{WFS}$ and $T_\text{DFS}$ denote the time costs for each WFS/DFS calculation with 128 parallel CPU processes based on the model 3rd Gen AMD EPYC™ CPUs (AMD EPYC 7763) in the Anvil Cluster at Purdue University.
In the remarks, BT stands for blind test, OSC standards for organic semiconductor, PAH stands for polynuclear aromatic hydrocarbons, Misc stands for miscellaneous group.}
\label{s1}
\vspace{3mm}
\\\hline
CSD\_entry & Space group & $Z^\prime$ & Rotors & ~Mass~ & ~~Success~~ & Energy Rank & $T_\text{WFS}$ & $T_\text{DFS}$ & Tier & Remarks\\
           & Symbol      &   &  /unit &        & rate (\%)   &             &  (min) &  (min) &      &\\
\hline
CYANAM01   & $Pbca$         & 1.00 & 0 &  42.0 & 27.9889 &      1/256      &  1&   1  & 1 &  Misc-cyanamide \\
BENZEN     & $Pbca$         & 0.50 & 0 &  78.1 & 27.5391 &      1/256      &  1&   1  & 1 & PAH-benzene\\
ADAMAN01   & $P$$2_1$       & 1.00 & 0 & 136.2 & 24.3164 &      1/256      &  1&   1  & 1 & Misc-adamantane\\
NAPHTA15   & $P$$2_1$/$c$   & 0.50 & 0 & 128.2 & 19.1406 &      1/256      &  1&   1  & 1 & PAH-naphthalene\\
ACEMID02   & $R$3$c$        & 1.00 & 0 &  59.1 & 17.2191 &      1/256      &  1&   1  & 1 & Misc-acetamide\\
BIPHEN     & $P$$2_1$/$a$   & 0.50 & 1 & 154.2 & 15.1367 &      1/256      &  1&   1  & 1 & PAH-biphenyl\\
AXOSOW01   & $Pbca$         & 1.00 & 1 &  56.1 & 10.3516 &     57/256     &  1&   1  & 1 & BT-XI\\
PHENAZ01   & $P$$2_1$/$a$   & 0.50 & 0 & 180.2 &  9.3750 &      1/256      &  1&   1  & 1 & Misc-phenazine\\
NICLAN     & $P$$2_1$       & 1.00 & 0 & 340.5 &  7.9102 &     39/256     &  1&   1  & 1 & OSC-DNTT\\
FORMAM     & $P$$2_1$/$c$   & 1.00 & 0 &  45.0 &  7.4219 &     54/256     &  1&   1  & 1 & Misc-formamide\\
OXALAC11   & $Pcab$         & 0.25 & 1 &  90.0 &  7.4219 &     22/256     &  1&   1  & 1 & Misc-oxalic acid\\
RESORA03   & $Pna$$2_1$     & 1.00 & 0 & 110.1 &  5.9570 &      9/256      &  1&   1  & 1 & Misc-resorcinol\\
MERQUY     & $P$$2_1$/$n$   & 0.50 & 0 & 334.5 &  5.8594 &     17/256     &  1&   1  & 1 & OSC-BDT\\
OXALAC02   & $Pcab$         & 0.50 & 1 &  90.0 &  5.4688 &     19/256     &  1&   1  & 1 & Misc-oxalic acid\\
AVIBEN     & $P$$2_1$       & 1.00 & 0 & 440.6 &  5.2094 &     21/256     &  1&   1  & 1 & OSC-DPhNDT\\
COUMAR01   & $Pca$$2_1$     & 1.00 & 0 & 146.1 &  5.0781 &      1/256      &  1&   1  & 1 & Misc-coumarin\\
SIRMIQ01   & $P$$2_1$/$n$   & 0.50 & 0 & 290.4 &  4.9020 &      1/256      &  1&   1  & 1 & OSC-BBBT\\
JAYDUI     & $P$$2_1$/$n$   & 1.00 & 0 &  44.1 &  4.0039 &      1/256      &  1&   1  & 1 & BT-VII\\
QUPHEN     & $P$$2_1$/$a$   & 0.50 & 3 & 306.4 &  3.6303 &      1/256      &  1&   1  & 1 & PAH-p-quaterphenyl\\
OKUPUG     & $P$$2_1$/$a$   & 0.50 & 2 & 392.5 &  3.5909 &      4/256      &  1&   1  & 1 & OSC-DPh-NDT\\
MERQOS     & $Pbca$         & 0.50 & 0 & 370.6 &  3.5156 &     15/256      &  1&   1  & 1 & OSC-BEDT-BDT\\
PAHYON01   & $C$2/$c$       & 1.00 & 0 & 100.1 &  3.4180 &      1/256      &  1&   1  & 1 & BT-VIII\\
UREAXX02   & $P$-42$_1$$m$  & 0.25 & 0 &  60.1 &  3.3933 &     30/256      &  1&   1  & 1 & Misc-urea\\
BOQQUT     & $P$2$_1$/$a$   & 1.00 & 0 & 153.2 &  3.2715 &     12/256      &  1&   1  & 1 & BT-IV\\
UVAHEF     & $P$2$_1$/$a$   & 0.50 & 2 & 392.5 &  3.0835 &      6/768      &  1&   1  & 1 & OSC-DPh-BTBT\\
CRYSEN     & $I$2/$c$       & 0.50 & 0 & 228.3 &  3.0570 &      1/768      &  1&   1  & 1 & PAH-chrysene\\
UVAGUU     & $P$2$_1$/$c$   & 1.00 & 2 & 392.5 &  2.7986 &      1/1024     &  1&   2  & 1 & OSC-Ph-NDT\\
MUVMIA     & $P$2$_1$/$n$   & 0.50 & 0 & 368.5 &  2.6348 &      1/256      &  1&   1  & 1 & OSC-3,10-DMeDNTT\\
JUFRIO     & $P$$2_1$/$c$   & 0.50 & 2 & 382.6 &  2.3788 &      1/768      &  1&   1  & 1 & OSC-MT-ADT\\
AFIGIH     & $P$-1          & 1.00 & 6 & 303.4 &  2.2327 &      1/768      &  1&   1  & 1 & Misc-thiophene deriv.\\
DBZCOR     & $C$2/$c$       & 0.50 & 0 & 400.5 &  2.1304 &      1/1280     &  1&   2  & 1 & PAH-dibenzocoronene\\
DUTGIK     & $P$$2_1$/$a$   & 0.50 & 2 & 604.8 &  2.1016 &      1/256      &  1&   1  & 1 & OSC-DPh-BBTNDT\\
ECENAD     & $P$$2_1$/$c$   & 0.50 & 0 & 402.6 &  1.9637 &     82/768      &  1&   1  & 1 & OSC-BNTBDT\\
XELYUJ     & $P$2$_1$       & 1.00 & 5 & 482.7 &  1.9458 &      1/2048     &  1&   4  & 1 & OSC-dithiophenyl deriv.\\
XULDUD     & $Pbca$         & 1.00 & 0 &  94.1 &  1.9317 &    422/768      &  1&   1  & 1 & BT-I\\
BZPHAN01   & $P2_12_12_1$   & 1.00 & 0 & 228.3 &  1.8069 &     48/1024     &  5&   2  & 1 & PAH-benzo(c)phenanthrene\\
MERQIM     & $P$$2_1$/$a$   & 0.50 & 0 & 334.5 &  1.7658 &     37/768      &  1&   1  & 1 & OSC-HTPBDT\\
VOBYAN     & $P$$2_1$/$n$   & 0.50 & 0 & 402.6 &  1.4879 &     53/768      &  1&   1  & 1 & OSC-anthracene-deriv. \\
SITJUC     & $P$$2_1$/$c$   & 0.50 & 2 & 600.8 &  1.4861 &     54/3054     &  6&   6  & 1 & OSC-IDBT-7d\\
HAJJIN     & $P$$2_1$       & 1.00 & 6 & 400.6 &  1.3669 &      1/1536     &  1&   2  & 1 & OSC-DEP-DTT \\
BETFOV     & $P$-1          & 1.00 & 0 & 274.4 &  1.3546 &     31/766      &  1&   3  & 1 & Misc\\
UVAHAB     & $P$$2_1$/$c$   & 0.50 & 2 & 392.5 &  1.3043 &     94/1280     &  4&   2  & 1 & OSC-DPh-NDT4 \\
WEXBOS     & $P$$2_1$/$c$   & 0.50 & 4 & 444.6 &  1.2810 &      3/256      &  3&   1  & 1 & OSC-thiophene oligomer\\
XULDUD01   & $P$$2_1$/$c$   & 1.00 & 0 &  94.1 &  1.2503 &    908/2304     &  6&   3  & 1 & BT-I\\
KUPWOJ     & $P$$2_1$/$c$   & 0.50 & 2 & 492.7 &  1.1189 &      2/2048     &  1&   4  & 1 & OSC-2,9DPhDNTT\\
MIVDEC     & $Pna$2$_1$     & 1.00 & 4 & 560.7 &  1.1186 &      4/3072     &  8&  15  & 1 & OSC-rubrene-deriv.\\
PYRENE07   & $P$$2_1$/$a$   & 1.00 & 0 & 202.3 &  1.0785 &    162/768      &  4&   1  & 1 & PAH-pyrene\\
TIDFES     & $P$-1          & 1.00 & 3 & 428.5 &  1.0664 &      9/1790     &  6&   7  & 1 & OSC-tetracene-deriv.\\
GUFJOG     & $P$$2_1$/$n$   & 1.00 & 0 & 125.1 &  0.9729 &      7/2304     &  1&   3  & 1 & BT-II\\
SUCXIZ     & $P22_12_1$     & 1.00 & 2 & 286.4 &  0.8888 &      1/2048     &  1&   3  & 1 & OSC-benzothiophene\\
KUPWUP     & $Pbca$         & 0.50 & 2 & 492.7 &  0.8155 &      1/2304     &  3&   5  & 1 & OSC-3,10-DPhDNTT\\
FUNZOE     & $P$2$_1$       & 1.00 & 4 & 268.3 &  0.8084 &     29/1024     &  4&   1  & 1 & Misc\\
ZZZDKE01   & $P$-1          & 1.00 & 0 & 328.4 &  0.7102 &     22/3324     &  2&   9  & 1 & PAH-hexacene\\
NACJAF     & $P$2$_1$/$n$   & 1.00 & 0 & 248.3 &  0.6925 &    415/3072     &  7&   4  & 1 & BT-XXII\\
DURNAH     & $P$2$_1$       & 1.00 & 4 & 372.5 &  0.6810 &      6/2048     &  1&   4  & 1 & OSC-Ph-BTBT-C4\\
DOCXEA     & $P$2$_1$/$n$   & 1.00 & 0 & 452.6 &  0.5930 &    152/2560     &  3&   6  & 1 & OSC-BBTNDT\\
MERRAF     & $P$2$_1$/$c$   & 0.50 & 0 & 342.6 &  0.4413 &    213/1536     &  9&   2  & 2 & OSC-BMBT-BDT\\
CEQGEL     & $P$-1          & 2.00 & 0 & 252.3 &  0.4096 &      1/3072     &  7&   8  & 2 & PAH-benzo(e)pyrene\\
FOJVAC     & $P$2$_1$       & 1.00 & 4 & 518.7 &  0.4053 &      9/2304     &  5&   5  & 2 & OSC-BBPh-NDT\\
AXIDER     & $P$2$_1$/$c$   & 0.50 & 4 & 604.6 &  0.3495 &     49/3840     & 10&   8  & 2 & OSC-rubrene-deriv.\\
FOJTUU     & $Pbca$         & 0.50 & 2 & 466.6 &  0.2839 &      1/4608     &  4&  10  & 2 & OSC-DNap-NDT\\
FIQSEG     & $P$2$_1$/$c$   & 0.50 & 2 & 282.5 &  0.2827 &   1963/4608     & 10&   7  & 2 & OSC-MT-BDT\\
ACSALA13   & $P$2$_1$/$c$   & 1.00 & 3 & 180.2 &  0.2777 &    261/16384    & 14&  26  & 2 & Misc-aspirin\\
WIDBAO     & $P$2$_1$/$c$   & 1.00 & 2 & 254.4 &  0.2651 &   1775/11008    &  5&  17  & 2 & BT-XIV\\
TBZPYR     & $Pn$2$_1$$m$   & 0.50 & 0 & 352.4 &  0.2413 &   2107/4864     & 25&  10  & 2 & PAH-b-tribenzopyrene\\
XATMOV     & $P$2$_1$/$c$   & 2.00 & 0 &  57.1 &  0.2224 &     86/4352     &  9&   7  & 2 & BT-XI\\
FLUANT     & $P$2$_1$/$n$   & 2.00 & 0 & 202.3 &  0.1855 &      1/54016    &  6& 165  & 2 & PAH-fluoranthene\\
QAXMEH53   & $P$-1          & 1.00 & 3 & 259.3 &  0.1841 &   1828/4352     & 28&   7  & 2 & Misc-ROY\\
OBEQOD     & $P$2$_1$/$c$   & 1.00 & 2 & 237.0 &  0.1755 &   4245/7936     & 16&  12  & 2 & BT-XVII\\
KEKJEQ     & $P$2$_1$/$n$   & 0.50 & 2 & 418.4 &  0.1642 &    230/6912     &  7&  13  & 2 & OSC-DP-NTCDI\\
ACSALA     & $P$2$_1$/$c$   & 1.00 & 3 & 180.2 &  0.1617 &    753/13312    &  9&  21  & 2 & Misc-aspirin\\
CAYKUJ     & $P$2$_1$       & 1.00 & 5 & 472.7 &  0.1601 &     23/15616    & 26&  34  & 2 & Misc.\\
XATJOT     & $Pca$2$_1$     & 2.00 & 3 & 246.2 &  0.1504 &     71/5376     & 54&   8  & 2 & BT-XIX-cocrysta\\
XAFQIH     & $P$-1          & 1.00 & 7 & 561.5 &  0.1486 &      9/24576    &  8&  64  & 2 & BT-XXVI\\
UJIRIO01   & $P$2$_1$/$c$   & 1.00 & 3 & 249.3 &  0.1406 &    560/16640    & 19&  27  & 2 & BT-VI\\
QUATER10   & $P$2$_1$/$a$   & 1.00 & 0 & 500.6 &  0.1311 &   1128/5374     & 36&  30  & 2 & PAH-quaterrylene\\
OBEQET     & $Pbca$         & 1.00 & 3 & 258.7 &  0.1227 &    166/6656     & 24&  10  & 2 & BT-XVIII\\
YIHVUI     & $P$2$_1$/$c$   & 0.50 & 1 & 490.7 &  0.1131 &   2729/11008    & 26&  21  & 2 & OSC-BTDT\\
OBEQUJ     & $Pbca$         & 1.00 & 0 & 120.1 &  0.1017 &  11461/12032    & 31&  17  & 2 & BT-XVI\\
UJIRIO05   & $Pbca$         & 1.00 & 3 & 249.3 &  0.0962 &    637/28160    & 14&  48  & 2 & BT-VI\\
HAMTIZ01   & $P$2$_1$/$n$   & 1.00 & 4 & 239.2 &  0.0957 &    478/14080    &  9&  24  & 2 & BT-X\\
XAFPAY01   & $P$-1          & 1.00 & 6 & 386.3 &  0.0917 &     43/13310    & 21&  32  & 2 & BT-XXIII\\
KONTIQ09   & $P$2$_1$/$c$   & 2.00 & 1 & 188.1 &  0.0910 &  1895/19456    & 19&  32  & 2 & BT-XXI-hydrate\\
ACRDIN05   & $Cc$           & 2.00 & 0 & 179.2 &  0.0897 &      1/18176    & 11&  36  & 2 & Misc-acridine\\
PYRZIN01   & $P$2$_1$/$c$   & 1.00 & 1 & 123.1 &  0.0770 &  50565/51200    & 28&  79  & 2 & Misc-pyrazinamide\\
QQQCIG04   & $Cmca$         & 0.25 & 4 & 532.7 &  0.0680 &    256/30976    & 44& 113  & 2 & OSC-ruberene\\
BOQQUT01   & $P$2$_1$/$c$   & 2.00 & 0 & 153.2 &  0.0597 &    602/42752    & 53&  82  & 2 & BT-IV\\
EVIDEV     & $P$2$_1$/$c$   & 0.50 & 2 & 344.6 &  0.0512 &   4788/13824    & 27&  23  & 2 & OSC-4TA-2\\
XAFQAZ     & $P$2$_1$/$c$   & 2.00 & 3 & 462.5 &  0.0501 &     13/113914   & 18& 284  & 2 & BT-XXV-cocrystal\\
JOHSOP     & $P$2$_1$       & 2.00 & 0 & 284.4 &  0.0251 &    914/19968    &265&  55  & 3 & OSC-ABT\\
WICZUF     & $P$2$_1$/$n$   & 2.00 & 1 & 245.3 &  0.0230 &  77460/80384    & 52& 155  & 3 & BT-XV\\
KONTIQ     & $P$2$_1$/$c$   & 2.00 & 1 & 188.1 &  0.0125 & 18876/128000   &230& 231  & 3 & BT-XXI-hydrate\\
LEVJON     & $P$2/$c$       & 0.50 & 4 & 696.5 &  0.0055 &    143/127992   &348& 383  & 3 & OSC\\
XAFPAY     & $P$2$_1$/$c$   & 1.00 & 6 & 386.3 &  0.0032 &     65/127958   &312& 286  & 3 & BT-XXIII\\
XAFPAY03   & $P$2$_1$/$n$   & 1.00 & 6 & 386.3 &  0.0013 &     53/127938   &312& 284  & 3 & BT-XXIII\\
OBEQIX     & $P$2$_1$/$n$   & 1.00 & 8 & 478.6 &  0.0009 &      5/127998   &362& 349  & 4 & BT-XX\\
XAFQON     & $P$2$_1$/$c$   & 3.00 & 4 & 201.7 &  0.0006 &  28062/127994   &219& 209  & 4 & BT-XXIV\\
YOKBIK     & $P$2$_1$/$a$   & 0.50 & 14& 464.8 &  0.0002 &    631/127992   &481& 384  & 4 & OSC-C8-BTBT\\
XAFPAY02   & $P$-1          & 2.00 & 12& 386.3 &  0.0000 &      0/127998   &367& 403  & 4 & BT-XXIII\\
UJIRIO02   & $P$2$_1$/$c$   & 2.00 & 6 & 249.3 &  0.0000 &      0/127998   &273& 297  & 4 & BT-VI\\
\hline
\end{longtable*}


\renewcommand{\thefigure}{A1}

\begin{sidewaysfigure}
    \vspace*{9cm} 
    \makebox[\textwidth]{ 
        \hspace*{-2cm} 
        \includegraphics[width=0.95\textwidth]{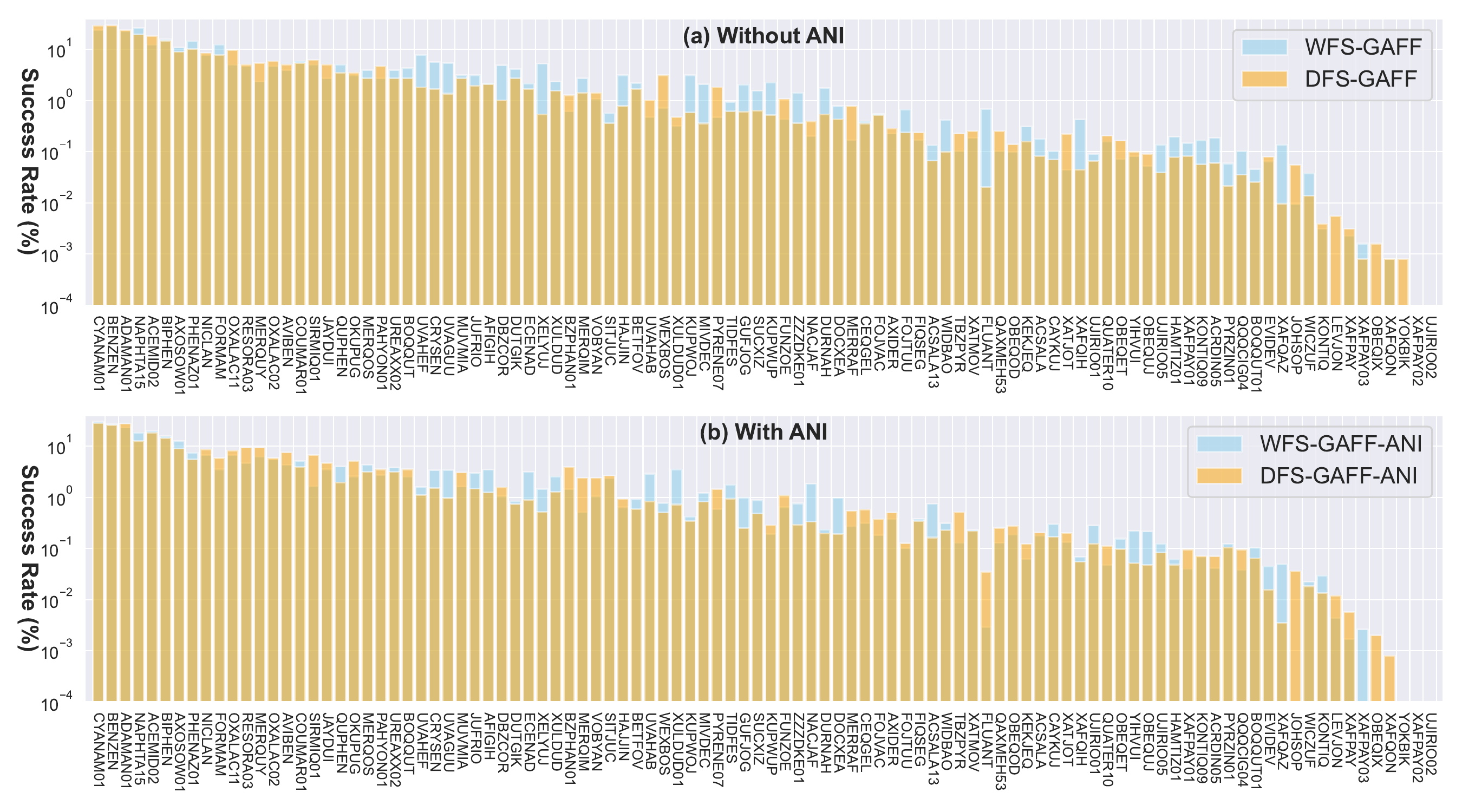}
    }
    \caption{The success rate distribution of 100 systems from four different sampling strategies.}
    \label{figs1}
\end{sidewaysfigure}



\end{document}